\documentclass[a4paper]{jpconf}
\usepackage{graphicx}
\usepackage{caption}
\usepackage{subcaption}
\usepackage{citesort}
\usepackage{xcolor}
\usepackage{booktabs}
\usepackage[acronym]{glossaries}

\newacronym{hpo}{HPO}{Hyperparameter Optimization}
\newacronym{hpc}{HPC}{High Performance Computing}
\newacronym{hep}{HEP}{High Energy Physics}

\begin{document}
\title{Hyperparameter optimization, quantum-assisted model performance prediction, and benchmarking of AI-based High Energy Physics workloads using HPC}

\author{Eric Wulff, Maria Girone, David Southwick, Juan Pablo García Amboage and Eduard Cuba}

\address{CERN, Esplanade des Particules 1, 1211 Geneva 23, Switzerland}
%\address{$^2$USC, Praza do Obradoiro, 0, 15705 Santiago de Compostela, A Coruña, Spain}
%\address{$^2$ETSE-USC, Rúa López de Marzoa, s/n, 15782 Santiago de Compostela, A Coruña, Spain}
% \address{$^3$University of Zurich, Rämistrasse 71, CH-8006 Zürich}

\ead{eric.wulff@cern.ch}

\begin{abstract}
% In the European Center of Excellence in Exascale Computing "Research on AI- and Simulation-Based Engineering at Exascale" (CoE RAISE), researchers from science and industry develop novel, scalable Artificial Intelligence technologies towards Exascale.
Training and \gls{hpo} of deep learning-based AI models are often compute resource intensive and calls for the use of large-scale distributed resources as well as scalable and resource efficient hyperparameter search algorithms. This work studies the potential of using model performance prediction to aid the \gls{hpo} process carried out on High Performance Computing systems. In addition, a quantum annealer is used to train the performance predictor and a method is proposed to overcome some of the problems derived from the current limitations in quantum systems as well as to increase the stability of solutions. This allows for achieving results on a quantum machine comparable to those obtained on a classical machine, showing how quantum computers could be integrated within classical ML tuning pipelines.

Furthermore, results are presented from the development of a containerized benchmark based on an AI-model for collision event reconstruction that allows us to compare and assess the suitability of different hardware accelerators for training deep neural networks.
\end{abstract}
\glsresetall

\section{Introduction}
% Training and hyperparameter optimization (\gls{hpo}) of DL-based AI models is often compute resource intensive and calls for the use of large-scale distributed resources as well as scalable and resource efficient \gls{hpo} algorithms \cite{Mlpf\gls{hpo}}.

Current state-of-the-art hyperparameter (HP) search algorithms such as Hyperband \cite{hyperband}, the Asynchronous Successive Halving Algorithm (ASHA) \cite{asha} and Bayesian Optimization Hyperband (BOHB) \cite{bohb} rely on a method of early termination, where badly performing trials are automatically terminated to free up compute resources for new trials to be started.

\begin{figure}[htb]
	\begin{minipage}{0.45\linewidth}
		\includegraphics[width=1\linewidth]{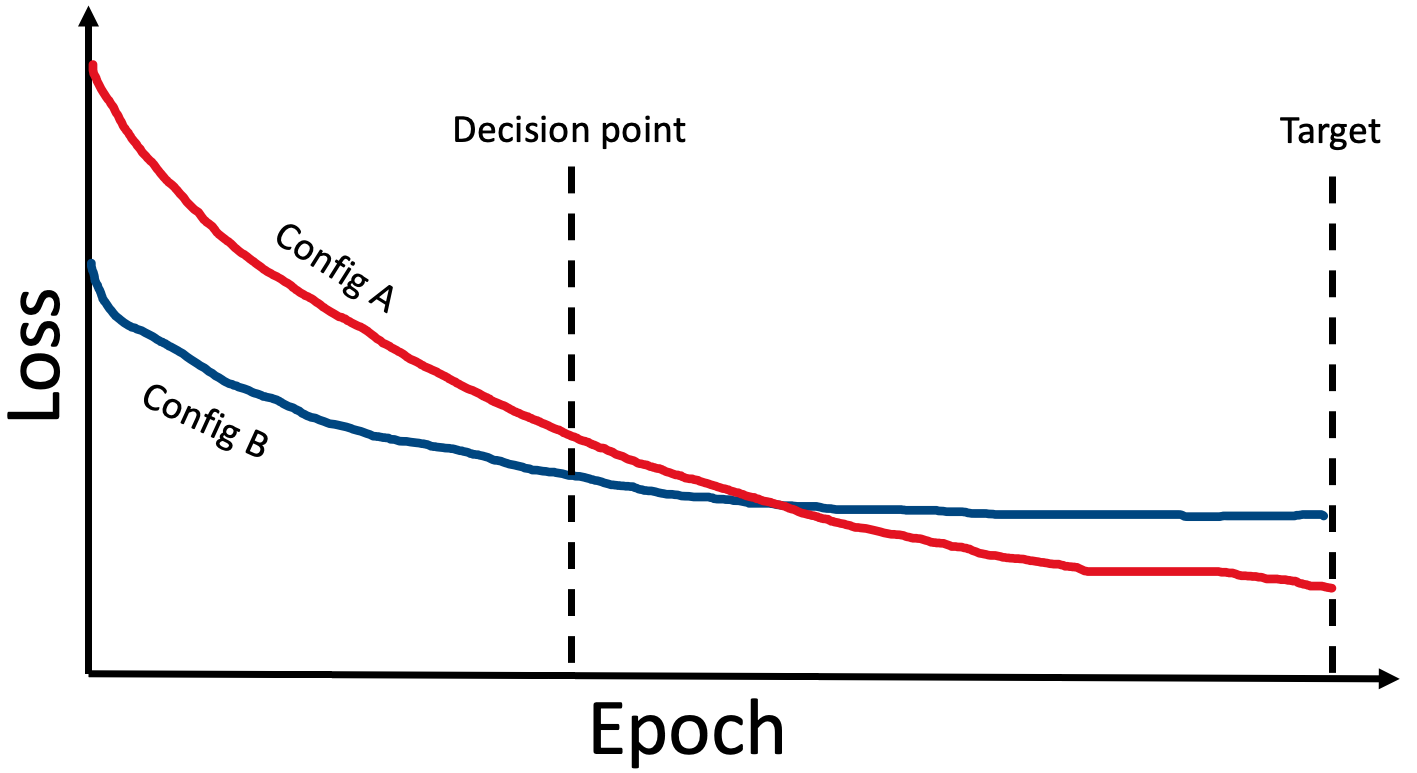}
		\\ \vspace{-7mm}
		\caption{Sketch of two learning curves resulting from different HP configurations.}
		\label{fig:sketch}
	\end{minipage}\hspace{1pc}%
	\begin{minipage}{0.55\linewidth}
		\includegraphics[width=.9\linewidth]{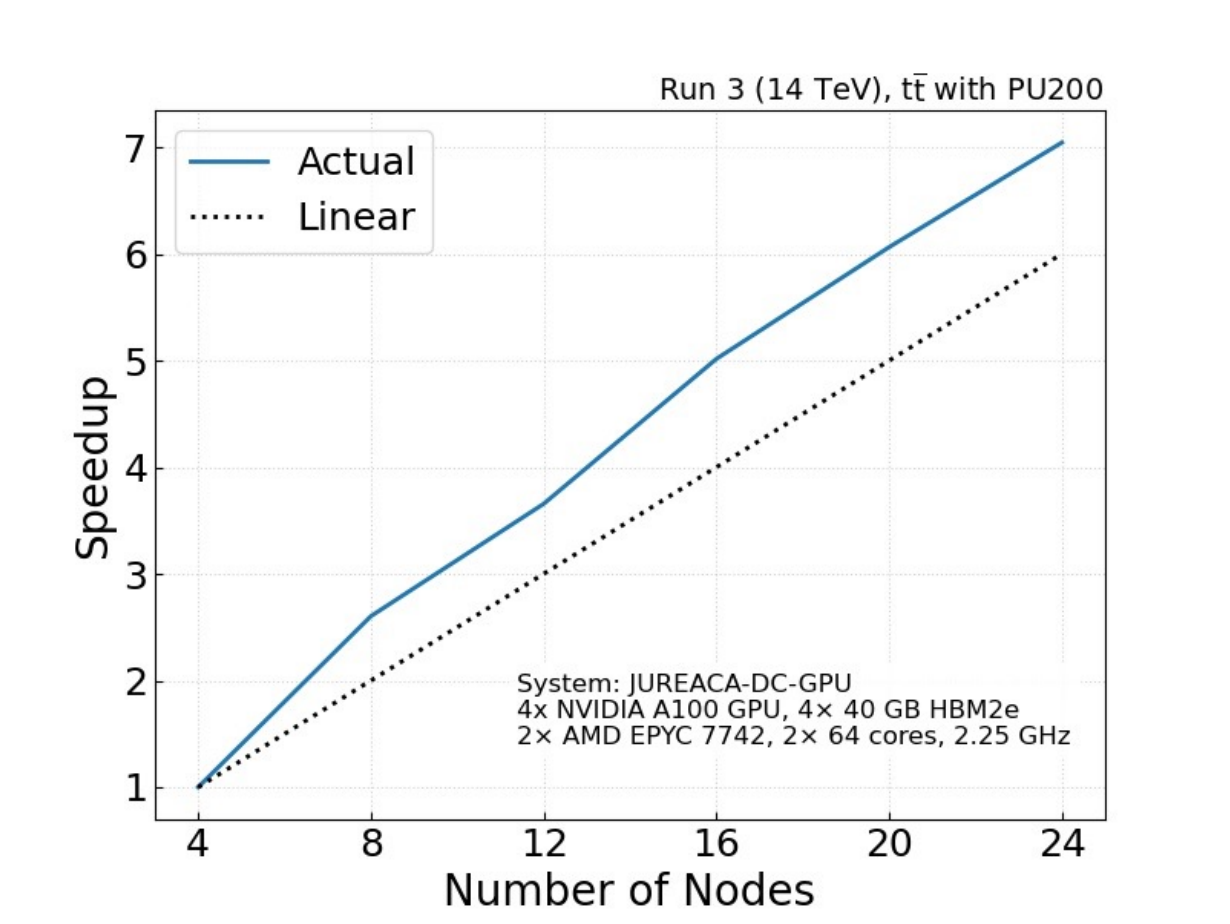}
		\\ \vspace{-7mm}
		\caption{Scaling of \gls{hpo} workflow on the JURECA-DC-GPU system.}
		\label{fig:hpo-scaling}
	\end{minipage} 
\end{figure}

Basing the stopping criterion on the relative ranking of trials according to a chosen metric, often accuracy or validation loss, can be problematic. Since the training process is non-linear, the ranking of trials at the decision point does not necessarily hold at the target point, see figure \ref{fig:sketch}. A potential solution to this problem is to use a non-linear stopping criterion, e.g., using Support Vector Regression (SVR), to predict the final model performance from a partially trained model \cite{perfpred}.

Large-scale \gls{hpc} systems are especially suited to run \gls{hpo} algorithms such as ASHA due to the superlinear scaling that can be achieved, as demonstrated in tests performed on the JURECA-DC \cite{jureca} system at the Jülich Supercomputer Centre (JSC), presented in figure \ref{fig:hpo-scaling}.

\section{Quantum support vector regression for model performance prediction}

The potential to speed up the \gls{hpo} process via performance prediction as well as the use of a quantum annealer (QA) to train the performance predictor is investigated. The Quantum-SVR (QSVR) performance achieved is comparable to classical SVR, showing that quantum computers are good candidates to be integrated within classical ML tuning workflows in the future.\\

\begin{figure}[htb]
	\begin{minipage}{0.5\linewidth}
		\includegraphics[width=1\linewidth]{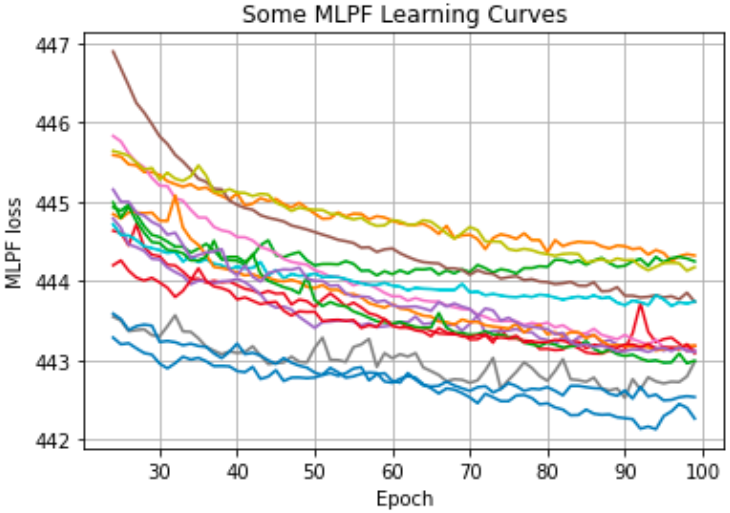}
		\\ \vspace{-7mm}
		\caption{The best learning curves from the learning curve dataset, zoomed in on epochs 25 to 100.}
		\label{fig:lc-dataset}
	\end{minipage}\hspace{1pc}%
	\begin{minipage}{0.5\linewidth}
		\includegraphics[width=1\linewidth]{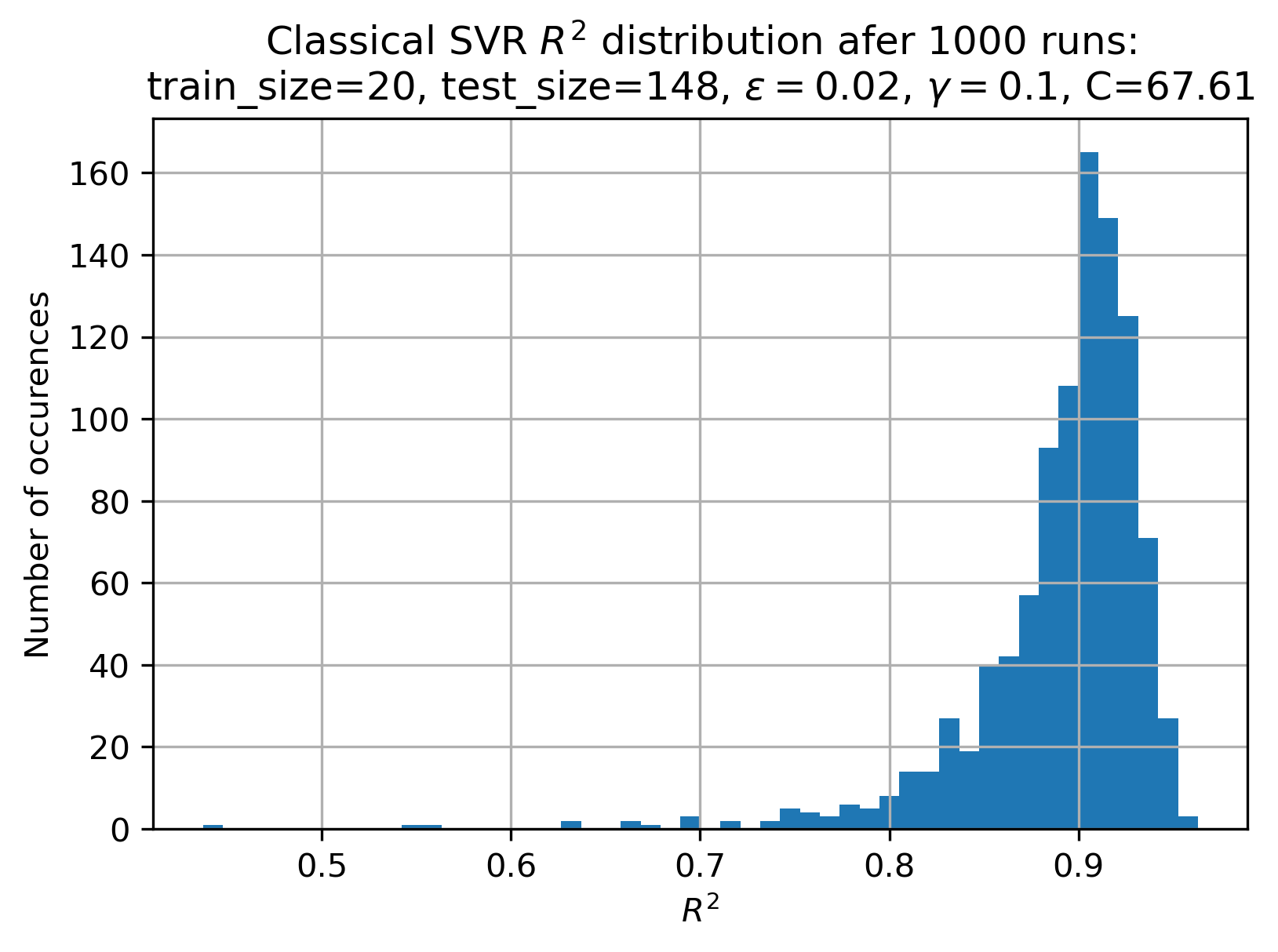}
		\\ \vspace{-7mm}
		\caption{Resulting $R^2$ distribution after training and testing an SVR on 1000 different splits on a classical computer.}
		\label{fig:r2-distribution}
	\end{minipage} 
\end{figure}

A Graph Neural Network (GNN)-based algorithm, developed for the task of machine learned particle flow (MLPF) reconstruction \cite{mlpf1,mlpf2} in \gls{hep}, acts as the base model for which studies are performed. A dataset consisting of learning curves and HP configurations was generated, see figure \ref{fig:lc-dataset}, by training 296 different configurations of MLPF on the publicly available Delphes dataset \cite{data_delphes}. Configurations were drawn randomly from the HP space defined in table \ref{tab:hp-space}. The trainings were run in a distributed data-parallel mode on compute nodes with four NVIDIA A100 GPUs each. Training one model for 100 epochs on such a node required roughly 16 hours of wall time. To speed up the generation of learning curves, trainings were run in parallel on 24 compute nodes on the JURECA-DC-GPU system at JSC.

% \begin{table}[h]
% \footnotesize
% \centering
% \begin{tabular}{@{}llllllll@{}}
% \toprule
%       & \begin{tabular}[c]{@{}l@{}}learning\\ rate\end{tabular}          & \begin{tabular}[c]{@{}l@{}}num graph\\ layers id\end{tabular} & \begin{tabular}[c]{@{}l@{}}num graph\\ layers reg\end{tabular} & dopout       & bin size                                                           & output\_dim                                                             & \begin{tabular}[c]{@{}l@{}}weight\\ decay\end{tabular}           \\ \midrule
% type  & \begin{tabular}[c]{@{}l@{}}log\\ uniform\end{tabular}            & \begin{tabular}[c]{@{}l@{}}quantized\\ uniform\end{tabular}   & \begin{tabular}[c]{@{}l@{}}quantized\\ uniform\end{tabular}    & uniform      & choice                                                             & choice                                                                  & \begin{tabular}[c]{@{}l@{}}log\\ uniform\end{tabular}            \\
% range & \begin{tabular}[c]{@{}l@{}}$(1e^{-6},$\\ $3e^{-2})$\end{tabular} & $[0, 4]$                                                      & $[0, 4]$                                                       & $(0.0, 0.5)$ & \begin{tabular}[c]{@{}l@{}}{[}8, 16, 32,\\ 64, 128{]}\end{tabular} & \begin{tabular}[c]{@{}l@{}}{[}8, 16, 32,\\ 64, 128, 256{]}\end{tabular} & \begin{tabular}[c]{@{}l@{}}$(1e^{-6},$\\ $1e^{-1})$\end{tabular} \\ \bottomrule
% \end{tabular}
% \caption{Hyperparameter space used to create the learning curve dataset.}
% \label{tab:hp-space}
% \end{table}

\begin{table}[h]
\footnotesize
    \centering
\begin{tabular}{ll}
    \br
    Hyperparameter                              & Sample distributions \\
    \mr
    Learning rate                     &  $\log{lr} \sim \textsc{U}[10^{-6}, 3\cdot10^{-2}]$ \\
    Classification graph layers                & \{0, 1, 2, 3, 4\} \\
    Regression graph layers           &  \{0, 1, 2, 3, 4\} \\
    dropout              &  $U[0.0, 0.5]$ \\
    bin size              &  \{8, 16, 32, 64, 128\} \\
    Output dimensions              &  \{8, 16, 32, 64, 128, 256\} \\
    Weight decay              &  $\log{wd} \sim \textsc{U}[10^{-6}, 10^{-1}]$ \\
    \br
\end{tabular}
\caption{Hyperparameter space used to create the learning curve dataset.}
\label{tab:hp-space}
\end{table}

To establish a baseline, classical SVR was studied by fitting 1000 models on different train/test splits with the same constraint on dataset size as required by the QSVR. Due to the limited number of qubits on the QA, only 20 training samples could be used for fitting. Results vary depending on the training split and statistics are shown in the figure \ref{fig:r2-distribution} and table \ref{tab:r2}.

\begin{figure}[h]
	\includegraphics[trim={0 0 0 11mm},clip,width=0.55\linewidth]{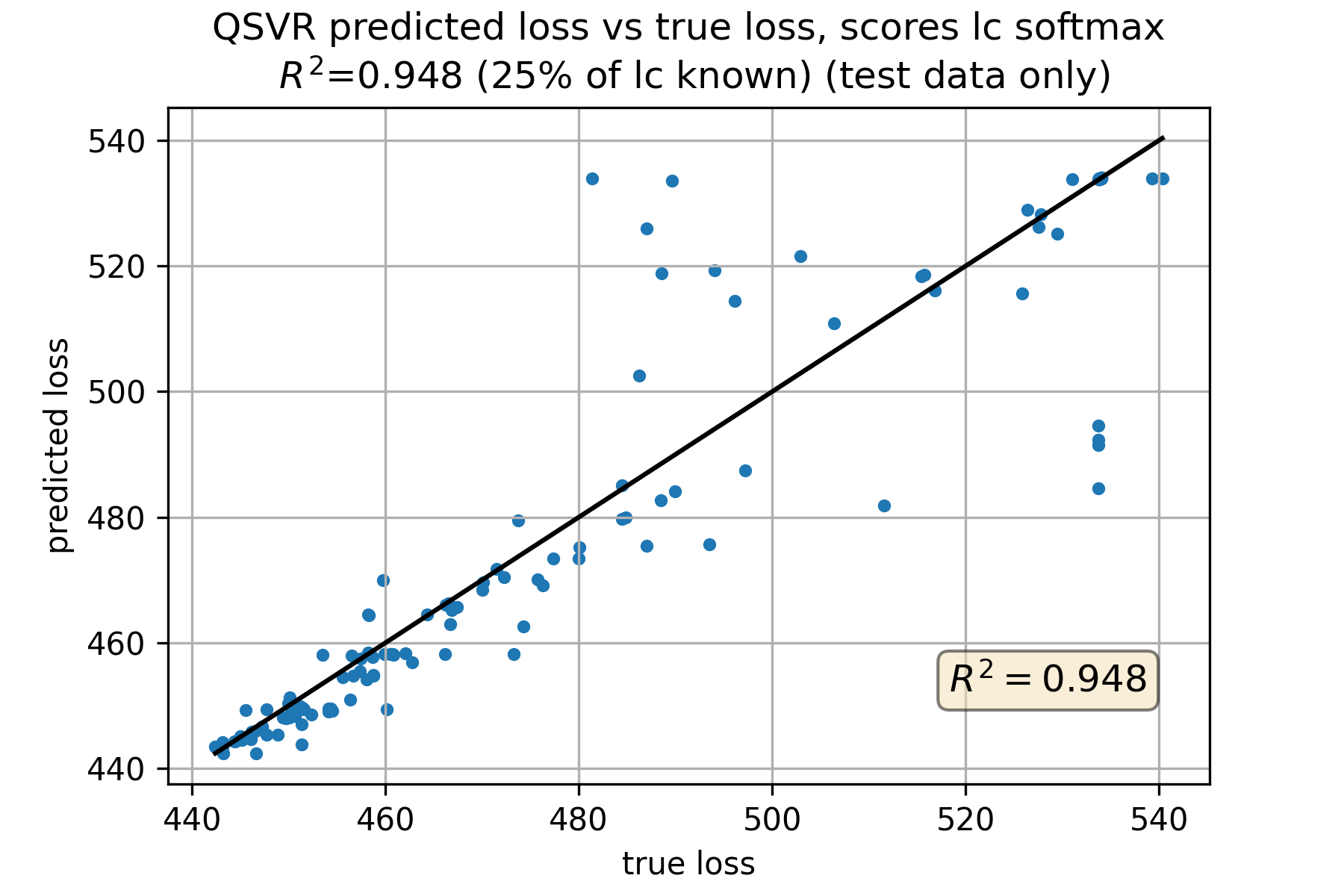} \hspace{2mm}
	\begin{minipage}[b]{0.28\linewidth} \caption{Predicted vs true validation loss values for the single best performing QSVR model. }
    \label{fig:best-single-qsvr}
	\end{minipage}
\end{figure}

Via the The European Center of Excellence in Exascale Computing "Research on AI- and Simulation-Based Engineering at Exascale" (CoE RAISE), access to JUPSI, a QA at JSC, was leveraged to train a QSVR model for model performance prediction. Due to the probabilistic nature of quantum processes the annealing is run multiple times and returns multiple solutions which can then be combined in different ways to create the final QSVR model. Different methods to combine the QSVR solutions are described in \cite{pasetto}. The predictions of the best performing QSVR are plotted against the true values in figure \ref{fig:best-single-qsvr}.

% \begin{figure}
%   \centering
%     \begin{minipage}[c]{0.6\textwidth}
%     \centering
%     \includegraphics[trim={0 0 0 12mm},clip,width=0.75\textwidth]{figs/scores_lc_softmax.png}
%   \end{minipage}
%   \begin{minipage}[c]{0.3\textwidth}
%     \centering
%     \begin{tabular}{|lr|}
% \hline
% \multicolumn{2}{|c|}{\textbf{$\mathbf{R^2}$ Summary}} \\ \hline
% \multicolumn{1}{|l|}{Maximum}                  & 0.948      \\ \hline
% \multicolumn{1}{|l|}{Minimum}                  & 0.742     \\ \hline
% \multicolumn{1}{|l|}{Mean}                     & 0.881     \\ \hline
% \multicolumn{1}{|l|}{Median}                   & 0.897      \\ \hline
% \multicolumn{1}{|l|}{Standard Deviation}       & 0.056      \\ \hline
% \end{tabular}
%   \end{minipage}
%   \caption{Predicted vs true validation loss values for the best performing QSVR as well as statistics for the $R^2$ score from 10 different training splits.}
% \label{fig:qsvr_comb}
% \end{figure}

% With the aim of stabilizing QSVR performance, as well as to increase the effective training set size, four separate QSVR models were trained on disjoint training sets of 20 samples each. Although this approach did not improve the maximum R$^2$ score, it did produce more stable results by significantly improving the worst performing split and reducing the standard deviation of R$^2$ scores between splits.

With the aim of stabilizing QSVR performance, different techniques were tested by combining several QSVRs trained on 20 samples each. The most successful approach found was to split the training set into disjoint subsets of 20 points and train a QSVR for each subset. The final ensemble prediction was calculated as the average of all the individual predictions.

\begin{figure}[h]
	\includegraphics[trim={0 0 0 11mm},clip,width=0.55\linewidth]{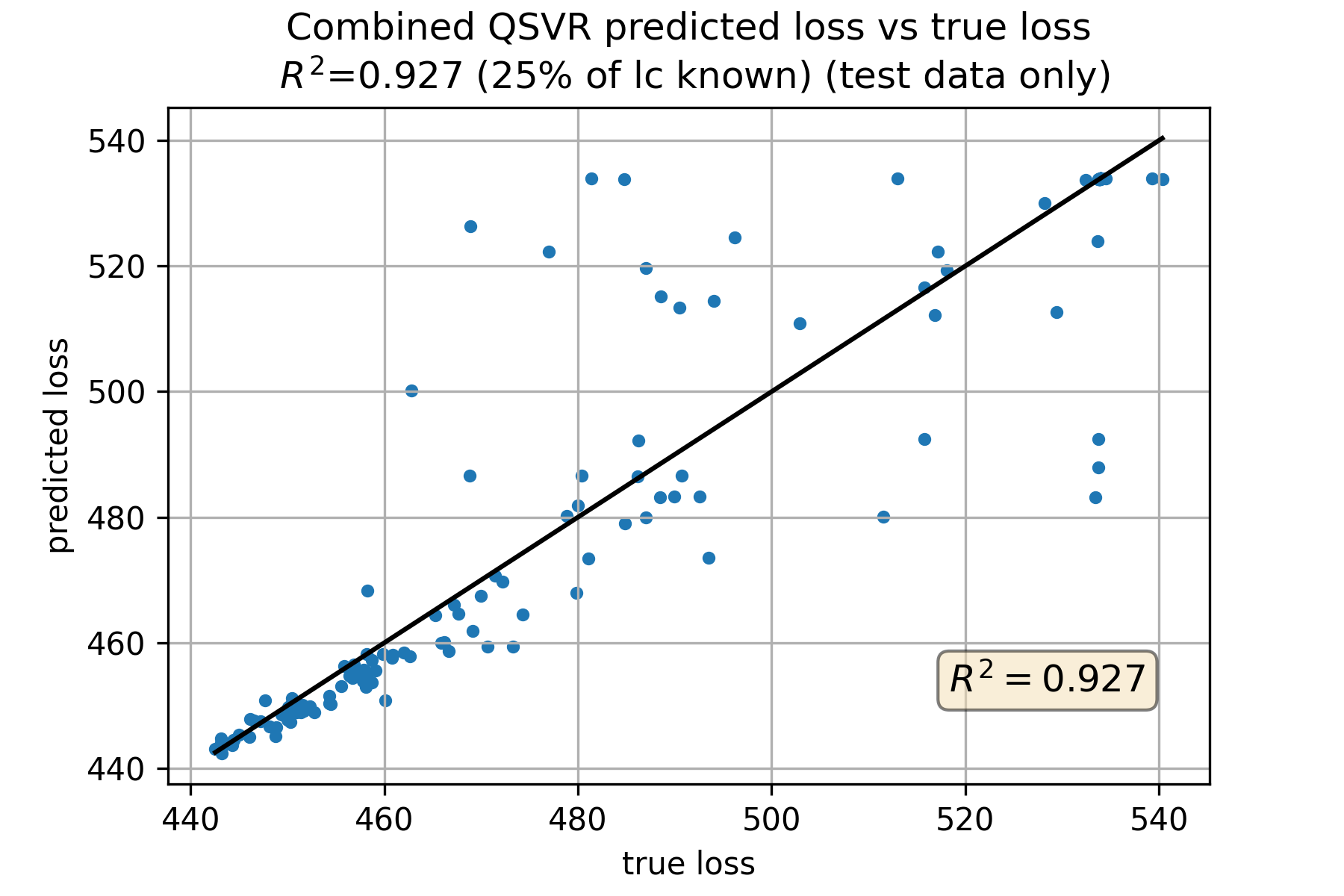} \hspace{2mm}
	\begin{minipage}[b]{0.45\linewidth} \caption{Best QSVR ensemble model predictions vs true values after running the QSVR combination technique 10 times for 10 different train/test splits using 80 training samples.}
		\label{fig:qsvr_comb}
	\end{minipage}
\end{figure}

To evaluate the QSVR combination technique, 80 training and 150 testing points were used. This experiment was repeated for 10 random train/test splits, where the results are presented in Figure \ref{fig:qsvr_comb}. Note that each prediction is made by combining 4 QSVRs trained on 20 points each. Although this approach did not improve the maximum $R^2$ score, it did produce more stable results by significantly improving the worst performing split and reducing the standard deviation of $R^2$ scores between splits, as can be seen in table \ref{tab:r2}. This is interesting as one of the problems faced while using the Q-SVR has been the instability of the results. Further improvement could be made by modifying the weights in the weighted average that combine the QSVRs, as currently a simple average is used. A comparison between the statistics of the $R^2$ scores for different models described in this work are presented in table \ref{tab:r2}.

% \begin{figure}
%   \centering
%     \begin{minipage}[c]{0.68\textwidth}
%     \centering
%     \includegraphics[width=0.75\textwidth]{figs/qsvr_true_vs_pred.png}
%   \end{minipage}
%   \begin{minipage}[c]{0.3\textwidth}
%     \centering
%     \begin{tabular}{|lr|}
% \hline
% \multicolumn{2}{|c|}{\textbf{$\mathbf{R^2}$ Summary}} \\ \hline
% \multicolumn{1}{|l|}{Maximum}                  & 0.927      \\ \hline
% \multicolumn{1}{|l|}{Minimum}                  & 0.857     \\ \hline
% \multicolumn{1}{|l|}{Mean}                     & 0.899     \\ \hline
% \multicolumn{1}{|l|}{Median}                   & 0.897      \\ \hline
% \multicolumn{1}{|l|}{Standard Deviation}       & 0.019      \\ \hline
% \end{tabular}
%   \end{minipage}
%   \caption{Results after running the Q-SVR combination technique 10 times for 10 different train/test splits using 80 training samples.}
% \label{fig:qsvr_comb}
% \end{figure}

% \begin{figure}[h]
% 	\includegraphics[trim={0 0 0 0},clip,width=0.5\linewidth]{figs/qsvr_true_vs_pred.png} \hspace{2mm}
% 	\begin{minipage}[b]{0.28\linewidth} \caption{Predicted vs true validation loss values for the best performing QSVR. The }
% 		\label{fig:asha-plots}
% 	\end{minipage}
% \end{figure}

\begin{table}[h]
\footnotesize
\centering
\begin{tabular}{lllllll}
\toprule
                    &       &       &       &       & Number of     \\
                    & Best  & Worst & Mean  & Std   & trainings     \\
\midrule
SVR           & 0.959 & 0.318 & 0.889 & 0.050 & 1000                \\
Sim-QSVR      & 0.949 & 0.383 & 0.901 & 0.045 & 100                 \\
QSVR          & 0.948 & 0.742 & 0.880 & 0.056 & 10                  \\
QSVR Ensemble & 0.927 & 0.857 & 0.899 & 0.019 & 10                  \\
\bottomrule
\end{tabular}
\caption{$R^2$ values and statistics for the different regression models. The SVR model is a classical SVR trained on a classical computer. The Sim-QSVR model is a QSVR model trained on a classical computer using simulated quantum annealing. The QSVR model is a QSVR trained on the QA and the QSVR Ensemble is the combined predictions from 4 QSVRs trained on the QA using 20 training samples each.}
\label{tab:r2}
\end{table}

\section{Containerized AI Benchmark}

\begin{figure}[h]
	\includegraphics[trim={0 0 0 14mm},clip,width=0.6\linewidth]{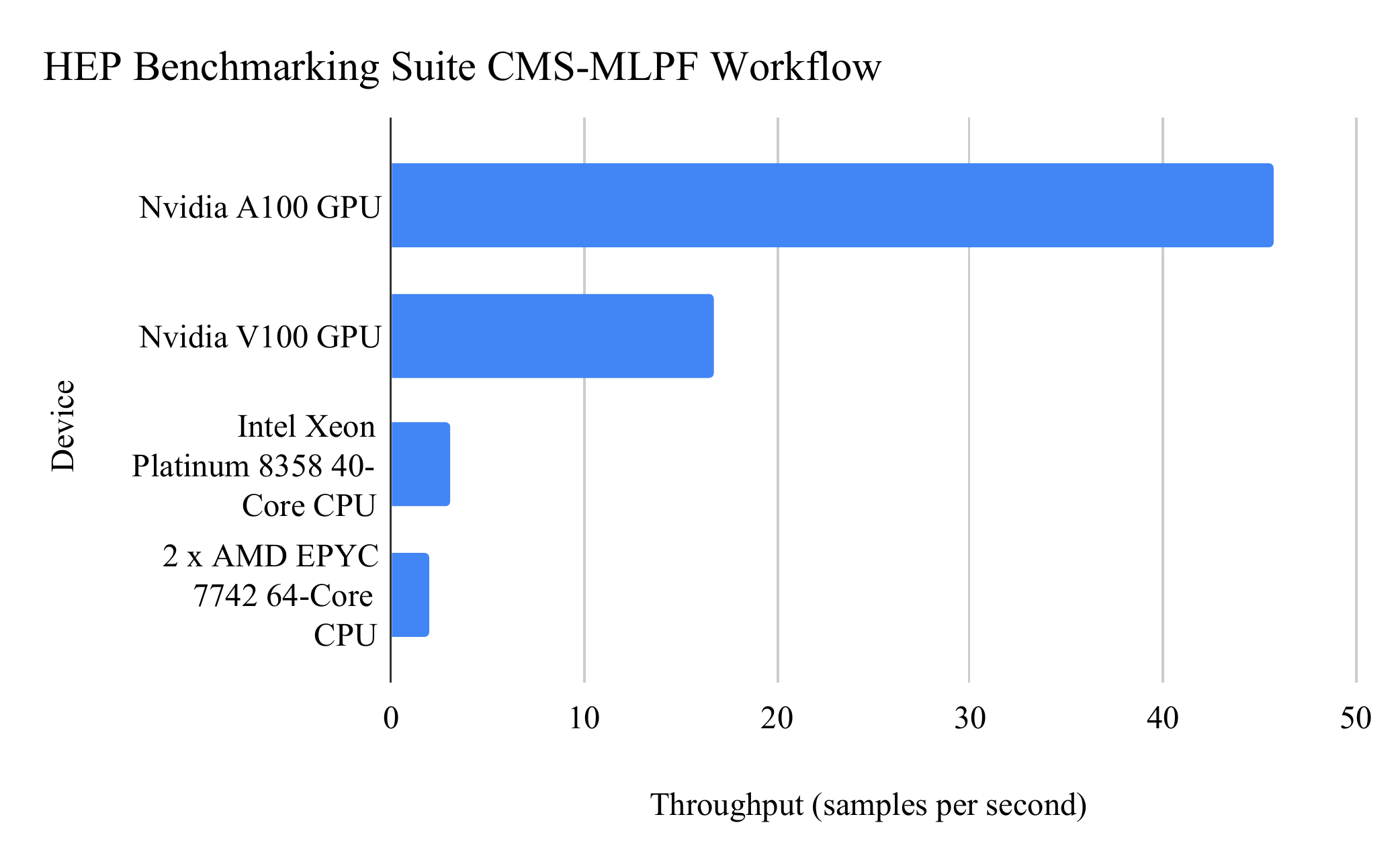} \hspace{2mm}
	\begin{minipage}[b]{0.28\linewidth}
    \caption{Benchmark results.}
    \label{fig:bmk}
    \end{minipage}
\end{figure}

The promise of better accuracy and reconstruction scalability at inference time for AI-based algorithms is preceded by resource-intensive training. A self-contained, reproducible and containerized benchmark based on the training of deep learning models is proposed to explore the feasibility of deploying AI-driven \gls{hep} applications to different \gls{hpc} environments. The metric shown in figure \ref{fig:bmk} is the training throughput, or training samples processed per second, on a subset of the publicly available Delphes dataset~\cite{data_delphes} in \gls{hep}score~\cite{hepscore2021} format.

\section{Conclusions}

\gls{hpc} systems are essential for running large-scale \gls{hpo} and distributed training and can significantly increase model performance as well as speed up the iteration of model development and training. It has been shown that superlinear scaling in speed-up can be achieved for \gls{hpo} workflows, indicating the benefits of \gls{hpc} for such use cases.

The strong potential of using performance prediction for \gls{hpo} was demonstrated, encouraging the use of this technique in future \gls{hpo} studies. It was also shown that, despite the current limitations of quantum computers, it is possible to train QSVR models on a QA while achieving prediction performance comparable to those obtained on a classical SVR. This encourages further studies in utilizing hybrid quantum/\gls{hpc} workflows for \gls{hpo} as well as in other use cases.

Finally, the development of a containerized benchmark application with an AI use case from \gls{hep} allows for quick and easy benchmarking of new hardware accelerators in the \gls{hep}Score format.

\ack
We thank our colleagues in CoE RAISE, in particular Andreas Lintermann and Marcel Aach for helpful discussions and feedback in the course of this work. We also thank our colleagues in the CMS Collaboration, especially Joosep Pata, Javier Duarte and Farouk Mokhtar for their collaboration on the MLPF studies.

Eric Wulff, David Southwick and Eduard Cuba was supported by CoE RAISE. The CoE RAISE project has received funding from the European Union’s Horizon 2020 – Research and Innovation Framework Programme H2020-INFRAEDI-2019-1 under grant agreement no. 951733.

The authors gratefully acknowledge the computing time granted through JARA on the supercomputer JURECA \cite{jureca} at Forschungszentrum Jülich. The authors gratefully acknowledge the Jülich Supercomputing Centre for funding this project by providing computing time through the Jülich UNified Infrastructure for Quantum computing (JUNIQ) on the D-Wave Advantage™ System JUPSI.

\begingroup
\section*{References}
\bibliography{references}
\endgroup

\end{document}